\documentclass[12pt]{article}
\input epsf.tex
\usepackage{cite}
\usepackage{latexsym}
\usepackage{graphicx}
\usepackage{amsmath,amstext}
\usepackage{comment}
\numberwithin{equation}{section}
\newcommand{\be}{\begin{equation}}
\newcommand{\ee}{\end{equation}}
\newcommand{\benn}{\begin{equation*}}
\newcommand{\eenn}{\end{equation*}}
\newcommand{\bea}{\begin{eqnarray}}
\newcommand{\eea}{\end{eqnarray}}
\newcommand{\bean}{\begin{eqnarray*}}
\newcommand{\eean}{\end{eqnarray*}}

\def\centeron#1#2{{\setbox0=\hbox{#1}\setbox1=\hbox{#2}\ifdim
\wd1>\wd0\kern.5\wd1\kern-.5\wd0\fi
\copy0\kern-.5\wd0\kern-.5\wd1\copy1\ifdim\wd0>\wd1
\kern.5\wd0\kern-.5\wd1\fi}}
\def\ltap{\;\centeron{\raise.35ex\hbox{$<$}}{\lower.65ex\hbox{$\sim$}}\;}
\def\gtap{\;\centeron{\raise.35ex\hbox{$>$}}{\lower.65ex\hbox{$\sim$}}\;}

\begin{document}
\begin{titlepage}
\begin{center}
\hfill SCIPP 08/07\\

\vskip 0.2in

{\Large \bf  Gauge Mediation with D terms}

\vskip 0.3in

Linda M. Carpenter,$^1$

\vskip 0.2in

\emph{$^1$ Santa Cruz Institute for Particle Physics,\\
     Santa Cruz CA 95064}

\begin{abstract}
I propose implementing General Gauge Mediation using the class of $SU(N) \times U(1)$ SUSY breaking models.   As an existence proof, I have utilized the 4-1 model in building multi-parameter gauge mediation.  These hidden sectors are relatively easy to use and avoid several model building pitfalls such as runaway directions.  In addition models require no special tuning and may produce as many parameters as general gauge mediation allows.
\end{abstract}

\end{center}
\end{titlepage}
\section{Introduction}

Minimal Gauge Mediation (MGM) provides a simple and predictive mass scheme
for supersymmetric models \cite{originalGM}.   MGM is achieved
by adding sets of vector like multiplets to the MSSM which are charged under
the normal gauge groups and couple to the hidden sector fields that participate in SUSY breaking.  These 'messengers' acquire both a supersymmetric and a nonsupersymmetric mass, and once they are integrated out, these terms give gauginos mass at one loop and scalars mass at two loops.  In its simplest incarnation, the messengers are coupled to a hidden sector singlet field  that acquires a scalar vev and F term in the hidden sector,
\be
 W= \phi X\overline{X} \rightarrow \langle \phi \rangle X\overline{X} + {\theta}^2 F_{\phi} X\overline{X}
 \ee
 where X are N messengers and $\phi$ the singlet field.
This yields one loop gaugino masses
 \be
 M_{\lambda i} =\frac{\alpha_i}{4\pi}\frac{F_{\phi}}{\langle \phi \rangle}
\ee
The entire mass spectrum is determined by a single
parameter ${F_{\phi}}/{\langle \phi \rangle}$ and we see that the gaugino and scalar mass
ratios are completely fixed.

However, because gauge mediation predicts a spectrum where sparticle
masses scale with powers of their gauge couplings, everything that
is charged under QCD is very heavy.  Thus even though this model is
predictive and flavor blind, problems persist in the spectrum.  In
particular the mass relationship $M_1$:$M_2$:$M_3$ $\sim$ 1:2:7 is
predicted for gauginos.  The chargino lower mass bound is 105 GeV
\cite{charginos}.  We might infer from indirect signal searches like
tri-leptons, that the lightest chargino is even heavier, approaching 150 GeV \cite{150GeV}.  In MGM setting the chargino above the mass bound then requires
a very heavy gluino due to the fixed gaugino mass ratio.  Lower bounds on the lightest charged sparticles
of 100 GeV also imply heavy squarks, if minimal gauge mediation
holds.  Squarks of 700 GeV would induce large corrections to the up
type Higgs mass parameter. The conditions for electroweak
breaking are known, and Higgs sector parameters must cancel down to
the Z mass.  Therefore the amount of tuning needed in the Higgs
sector is of order $(m_{h_u}/mz)^2$, or sub percent.

A compressed and lighter spectrum would alleviate tuning problems
that exist in the Higgs sector and open up SUSY parameter space to
new and interesting signals. This would require modifying the gauge
mediated predictions for sparticles.  Meade et al. have laid out the
formalism of General Gauge Mediation whereby the gauge mediated
spectrum may be determined by up to 6 parameters, including 3
independent parameters in the gaugino sector\cite{GGM}. Several
recent models fall under the category of GGM model building, with
weakly coupled renormalizable operators employing chiral fields
only. For example Extraordinary GM compresses the spectrum without
altering the gaugino mass prediction of MGM \cite{implGGM}. Other
proposals compress the spectrum and achieve the full range of GGM
parameters \cite{implGGM2}.

Weakly coupled renormalizable models which change the
gaugino mass ratio prediction require, at least, splitting the
doublet and triplet messenger couplings and coupling a single set of
messengers to multiple scalars. Thus we would have a superpotential
like, \be W = (\lambda_1 \phi_1 + \lambda_2 \phi_2 +... \lambda_i
\phi_i)X\overline{X} \rightarrow \sum \lambda_{qi} \phi_i
q\overline{q}+\sum \lambda_{li} \phi_i\l\overline{l} \ee Now we may
make the following field redefinitions:

\be Z \equiv (\sum \lambda_{qi} \phi_i)  ;  Y\equiv (\sum
\lambda_{li} \phi_i) \ee
Gaugino masses are now proportional to two
scales;
\be m_3 =\frac{\alpha_3}{4\pi} \frac{F_Z}{\langle z \rangle}, m_2
=\frac{\alpha_2}{4\pi}\frac{F_Y}{\langle y \rangle},
m_1=\frac{\alpha_1}{4\pi}(\frac{F_Y}{\langle y \rangle}+2/3\frac{F_Z}{\langle z \rangle})
\ee

We
see that the gaugino mass ratio of minimal gauge mediation is not preserved and we have achieved a two parameter model.

Models may be complicated even further by adding multiple scalars
and multiple messengers.  However models like this pose difficult
model building challenges.  For example, in models with multiple
messengers,  hypercharge D terms induce one loop masses for scalars
proportional to their hypercharge unless an interchange symmetry of
the messengers can be made to appear in the low energy theory.  In
models with multiple scalars which are built purely out of chiral
fields, some care is required to make sure the theory is stabilized
far from runaway directions so that all fields acquire proper vevs.
In addition there is a generic problem with phases.  In minimal
gauge mediation, gaugino and scalar masses all come from a single
mass scale, there are no relative phases between the gaugino masses.
However, when model building with multiple scalars and messengers,
splitting the gaugino mass requires the addition of many new
couplings and in general phases occur.

Instead, I propose the introduction of a single new
source of SUSY breaking from a hidden sector $U(1)$. This generates
a new operator in the theory, a non-supersymmetric
mass term which can be added to alter the minimal
gauge mediation prediction without multiple scalars or multiple
messengers.  The GGM parameter counting is distinctly different from
models in \cite{implGGM2}.  In addition, the hidden sector dynamics
can be implemented in simple and familiar $SU(N) \times U(1)$ models.  In
Section 2 I introduce the D-term operator and use it to build the
simplest GGM model.  In section 3 I review the dynamics of the 4-1
hidden sector. In section 4 I use other operators in the 4-1 model to
build gauge mediation and make an attempt at a unified model without
phases. Section 5 concludes.

\section{SUSY Breaking D terms}
In addition to F terms in the hidden sector, we may consider another
source of SUSY breaking,  a U(1) gauge field whose $D$ term acquires
a vev by some dynamical mechanism.  Since we want a $D$ term that is
 the same size as the overall SUSY breaking scale, we may deduce that
the $D$ term vev is itself closely connected to, even required for,
supersymmetry breaking.  The lowest dimension new operator
that one may write down with all indices contracted has the form

\be \frac{c}{{M}^2} \int d^2 \theta W^{'} W^{'}  X \overline{X} ,\ee where X is matter in a vector-like representation.  When the $D$
term is set to its vev this term becomes \be
 {c}\frac{D^2}{M^2} x\overline{x}
\ee   This is an additional $B$ term, which is a
source for nonsupersymmetric masses.  Such a term has been used as a source for SUSY breaking messenger masses for example in \cite{Dterms}.  The new operator only adds one
more parameter to the low energy theory, the scale $\sqrt{c}D/M$,
so we may maintain an economy of parameters.

Scalar masses for squarks and sleptons cannot be generated through
direct contact terms with the hidden sector gauge field.  Holomorphy
prevents us from writing such a term in the superpotential.  Instead
the lowest dimension mass term we may write is $\frac{1}{M^6}\int
d^4 \theta W^{'}W^{'}W^{'\dagger}W^{'\dagger}QQ^{\dagger}$, which is
highly suppressed and not generated by any divergent diagrams.

\subsection{A Simple Way To Use D-terms}

Consider a messenger superpotential with a single scalar field Z
that gets an F term and a scalar vev, and a hidden sector U(1)
field. \be W = fZ+ y_Q ZQ\overline{Q}+y_L Zl\overline{l} +
\frac{\lambda_Q}{M^2} W^{'}W^{'}Q\overline{Q}+\frac{\lambda_L}{M^2}
W^{'}W^{'}l\overline{l} \ee  where the couplings for doublet
and triplet messengers have been split.  Couplings between scalars fields
and the extra gauge fields may be forbidden by R symmetry.

\noindent Z gets an F term and a scalar vev $Z = \langle z\rangle + \theta^2 F_z$.
Messengers get a SUSY breaking mass from the F term and the extra D
term vev. Define $B= D^2/M $ so the gaugino masses are
\be
m_3=\frac{\alpha_3}{4 \pi}  (\frac{F_z}{z} + \frac{\lambda_Q B}{ y_Q z}),
m_2=\frac{\alpha_1}{4 \pi}  (\frac{F_z}{z} + \frac{\lambda_L B}{y_L z})
\ee The B term may be chosen to be of the same order as $F_z$.  If the ratio of
couplings $\lambda_Q/y_Q$ is smaller than the ratio of
$\lambda_L/y_L$ we lower the mass ratio of gluinos to the
other MSSM gauginos.  Notice that there are three distinct
parameters $F_z/z$, $\lambda_Q B/ y_Q z$, and $\lambda_L B/ y_L z$.

\section{The $4-1$ Model}

We now must address the best way to achieve a D-term vev.  To get a
D term of sufficient size, comparable to the overall scale of SUSY
breaking, we may build a model in which the $U(1)$ is required for
supersymmetry breaking.  The '4-1' Model of Dine and Nelson is a
simple and interesting example \cite{4-1}.

The model has an $SU(4)\times U(1)$ gauge group. The matter content
is as follows (subscripts indicate $U(1)$ charges): an antisymmetric
tensor $A_2$, a fundamental $F_{-3}$, an anti-fundamental
$\bar{F}_{-1}$ and a singlet $S_4$.  There is only one allowed
superpotential term, \be W=\lambda S_4 F_{-3}\bar{F}_{-1} \ee
 $SU(4)$ then confines and the gauginos condense generating a
non-perturbative term in the superpotential,
\be
W=\lambda S_4 F_{-3}\bar{F}_{-1} + \frac{\Lambda_4^5}{\left(\bar{F}_i
    F^jA^{ik}A^{lm}\epsilon_{jklm}\right)^{1/2}}.
\ee
 This model contains a non-anomalous R symmetry which is broken
 once the cosmological constant is tuned to zero, and hence a massive
 R axion \cite{Raxion}. The scale of SUSY breaking we will assume is
 high enough that the R axion is unobservable.

Making the choice, \be A_2=\left(\begin{matrix} a\sigma_2 & \ \\ \ &
a\sigma_2\end{matrix}\right), F=\bar{F}=\left(\begin{matrix}
b\\0\\0\\0\end{matrix}\right), S=c. \ee With the rescaling,
$\phi\rightarrow \frac{\Lambda}{\lambda^{1/5}} \phi$,  the D-term is
\be D_1=g_1\frac{\Lambda^2}{\lambda^{2/5}}(2|a|^2-4|b|^2+4|c|^2) \ee
the scalar potential F-term contribution is, \be
V_F=\lambda^{6/5}\Lambda^4 \left(|b|^4+\left|2 b c
-\frac{1}{ab^2}\right|^2 + \left|\frac{1}{a^2 b}\right|^2\right) \ee
We may now minimize the potential.  Notice that without the D-term
there is a runaway direction.

We may take $b \sim \epsilon$ for $\epsilon$ arbitrarily
small while $a \sim 1/\epsilon$ and $c \sim 1/\epsilon^2$.  Here
we can solve all of the F term equations. As we go out in the
runaway direction SUSY is restored.  However, as we turn on the
coupling $g_1$ we find we can no longer satisfy the F and D term equations
and SUSY is broken everywhere.  To avoid running away to a
supersymmetric minimum, we must generate a D term.  The term $D^2$ is
quartic in fields and for very small $g_1$ the minimum is far from the
origin.  Because of quartic behavior, as we turn $g_1$ up, the
minimum moves in closer to zero and the D term becomes small compared to the F term.  Note that The F term is always larger than the D term but regions of parameter space exist, for $\lambda \sim 10 g_1$, where they are of the same order.  We will see later how the size of this ratio effects phenomenology.

In addition to generating a D term for the
U(1), the 4-1 model also gives an additional useful operator for
model building, the gaugino condensate of the $SU(4)$ gauge
multiplet.

\section{The Gaugino Condensate}

We see that the in the 4- 1 model, in addition to having a $U(1)$ D
term, there is also a gaugino condensate.  Gaugino condensates are useful
for generating $\mu$ terms, see for example \cite{WW}.  Proceeding in a
way similar to the previous section, we see that we can couple
messengers to the gaugino condensate as well as to the D terms.  We
write the messenger superpotential

\be W=
y_1\frac{W_1W_1}{M^2}X\overline{X}+y_4\frac{W_4W_4}{M^2}X\overline{X}
\ee

There is now a B-term for the scalar messengers as well as $mu$
term generated by gaugino condensation.

\be
B=y_1 D^2/M^2;   \mu = y_4 \Lambda^3/M^2
\ee

We have built the operators needed for gauge mediation not out F
terms and vevs of chiral fields, but from gauge D terms and gaugino
condensates.  Gaugino masses are proportional to the ratio of B and
$\mu$ \be M_{\lambda}\sim y_1D^2/y_4\Lambda^3 \ee and are not
dependent on the scale M.  This simple model does not break
the gaugino mass ratio prediction of MGM, but instead reproduces the minimal gauge mediated
phenomenology.  Achieving the multiple parameters of GGM once again requires splitting
the messenger couplings.  Below, the messenger sector consists of a single set of messengers in the $5$, $\overline{5}$ representation however one may imagine repeating these steps for multiple sets of messengers in $5$, $\overline{5}$ or $10$, $\overline{10}$ representations.
 \be W=
y_1\frac{W_1W_1}{M^2}Q\overline{Q}+y_4\frac{W_4W_4}{M^2}Q\overline{Q}+l_1\frac{W_1W_1}{M^2}l\overline{l}+l_4\frac{W_4W_4}{M^2}l\overline{l}
\ee

Writing everything in terms of $\Lambda$ we have the relation

\be M_3 =
\frac{\alpha_i}{4\pi}\frac{y_1B}{y_4\mu}=\frac{\alpha_i}{4\pi}\frac{y_1g_i^2\Lambda(2|a|^2-4|b|^2+4|c|^2)^2
}{y_4 \lambda^{4/5}} \ee
\be M_2 =
\frac{\alpha_i}{4\pi}\frac{l_1B}{l_4\mu}=\frac{\alpha_i}{4\pi}\frac{y_1g_i^2\Lambda(2|a|^2-4|b|^2+4|c|^2)^2
}{y_4 \lambda^{4/5}} \ee In general $y_1$, $y_4$, $l_1$, $l_4$ may all be
different from each other.  What we need to break the MGM gaugino
mass prediction is that ${y_1}/{y_4}$ not be equal to
${l_1}/{l_4}$.  In order to avoid messenger vevs we must have $B< \mu^2$ or

\be
 \frac{g_1^2}{\lambda^{4/5}} (2|a|^2-4|b|^2+4|c|^2)^2  <  \Lambda^2/M^2
\ee
For the correct spectrum we may pick point like $\Lambda \sim 10^{8}$, M $\sim 20\Lambda$
 with couplings, $\lambda = 2.6$x$10^{-2}$ and $ g_1 = 6$x$10^{-1}$ and y's and l's of order $10^{-1}$.  We get a spectrum with gauginos in the hundred GeV range and no vevs for messengers.  Since there are two independent parameters for gluinos and winos, we may expect a spectrum with light squarks without the need to tune couplings.  This model achieves 2 parameters of the possible 6 of GGM.  If we had chosen messengers in a $10$, $\overline{10}$ representation we would have gotten a three parameter spectrum.  In fact, using the 4-1 models, the predictions for number of parameters and low energy spectrum follow from those in \cite{implGGM2} where our scales are set by D terms and gaugino condensates rather that F terms and vevs of chiral fields.

\subsection{Extra Operators}

We may now attempt to write down potentially dangerous operators
that get generated in the Kahler potential.  The most important is a coupling
of hidden sector fields and messenger fields, \be K=\lambda \int d^4\theta
\frac{F^{\dagger}FX^{\dagger}X}{M^2} \ee This operator does not
break R symmetry and cannot contribute to gaugino masses. However it
will induce an operator which is another source for scalar
masses, and has been well studied in \cite{hiddensector}.

This is an extra mass term for messengers; and since messengers only couple to MSSM fields with SM gauge couplings, this new contribution will yield flavor blind masses.  However these are not the standard mass terms of minimal gauge meditation.  The scalar mass contribution from the new operators is

\be
m^2_i = - f\sum_a\frac{g_a^4}{128 \pi^4} S_Q C_{ai} {\rm Str} M_{mess}^2 \log(\frac{M^2}{\Lambda^2})
\ee
where S is the Dynkin index of the messengers and $C_{ai}$ is the Casimir for the scalars.  Notice that unlike the standard GM contribution, there is running from the scale of the cut off M, presumably where we have integrated out some heavy fields to generate the operators $WWX\overline{X}$, to the scale at which the 4-1 model gauginos condense hence the log factor.  This is scaled by powers of this operators anomalous dimension.  In addition, this operator will be down by a factor of $\alpha_4(M)$ compared to the standard GM scalar mass contribution since this operator involves two insertions of the hidden sector F terms.  As long as the F terms are of manageable size and appear with a reasonable coefficient, we expect this operator not to dominate or drastically alter the spectrum.  However, if the F terms become large and the log does not scale away with large negative anomalous dimensions this contribution can become as important as the standard GM contribution to scalar masses or even dominant.  If the sign of the operator is negative the spectrum may even become tachyonic.  The trick then is to stay in regions of parameter space where F terms are not too large.  Another way around this constraint would be to forbid such operators all together.  For example, if the hidden sector fields were sequestered from MSSM fields using boundary conditions in 5-D, these extra contributions may be extremely suppressed.

\subsection{A Unified Model}
 In minimal gauge
mediation we avoid relative phases in the gaugino sector because all
gaugino masses come from a single mass parameter.  However models
with split gaugino masses usually have a relative phase. We would like as few phases as possible.  In addition we would like
to make a model that is as simple as possible.  We might try to begin with a
unified SU(5), then break it to $SU(4)\times U(1)$ needed for 4-1 SUSY breaking.  A $10$ and
one $\overline{5}$ of $SU(5)$ provide all of the chiral fields needed
for the 4-1 model.  In general we would begin with different
couplings of the $W$'s to doublet and triplet messengers.  The operators $WWX\overline{X}$ are generated by integrating out fields carrying quantum numbers under $SU(4)$ and $U(1)$.  If the gauge fields are unified at some high energy we expect that once SU(5) breaks, the $W_1$ and $W_4$ messenger couplings will split.  However it may be possible that the relative phases between terms-which start off the same when SU(5) is unified- remain the same.  This depends on the dynamics at the high scale.  What follows is an attempt to build a model where $SU(4)$ and $U(1)$ unify.

After some numerical estimates, we find that we may get the
correct order of magnitude for the MSSM field masses and avoid
tachyonic messenger masses if the gaugino condensation scale
$\Lambda$ is only a few decades above the cut-off $M$.

We may compute the scale at which SU(4) confines
by finding the pole in \be g(\mu)= \frac{g(\Lambda)}{1+\frac{b
g(\Lambda)^2 ln(\frac{\Lambda}{\mu})}{8\pi^2}} \ee where b for
$SU(N)$ is $3N_c - N_f$ here 8. 

Run this coupling up to the unification scale, and it is the value of the coupling $g_1$ at high energy which will
then run down.

We see that if we run over two decades, our unified coupling is
$g(\Lambda) \sim 1.4$.  The difficulty with this scenario is that for
running over only a few decades the U(1) and SU(4) couplings do not
split very much, it is not possible to make $g_1$ small.  Therefore,
the minimum of the potential comes close to the origin and the F
terms are generically much bigger than the D terms.  In the non-unified model one is free to pick smaller values for $g_1$ and this was not as great of a concern.

The scalar masses are now dominated by the contribution mentioned in the previous section.  Unless the hidden sector fields are sequestered with extra dimensions, or the extra operator has very large negative anomalous dimensions the extra contribution to scalar masses will be of the same order as the standard gauge mediated contribution.  If the sign of these contributions is negative, some scalars may become tachyonic.  In addition the extra contributions may reintroduce tuning by increasing scalar masses.  This is not a concern if a suitable sequestering mechanism can be found.  Even before we worry about finding suitably high energy dynamics, the viability this model is in question.   Thus model building without gaugino phases requires further study.

\section{Conclusions}
It possible to build simple implementations of GGM by stepping outside the bounds of weakly coupled
chiral models.  Here I have demonstrated the viability of using the
4-1 SUSY breaking model.  However it is likely that a range of
$SU(N) \times U(1)$ models may also yield good results.  In addition, I
have shown the existence of a mixture of F-term and 4-1 style SUSY
breaking that has different parameter counting than previous
attempts at GGM completions.  Hence I have generated new models for which a compressed SUSY spectrum is possible.  Attempts to build models without phases in the gaugino sector fare worse.  The minimal unified attempt to build
models leads to large contributions to scalar masses.  The problem of phases therefore is not resolved unless a suitable sequestering mechanism can be implemented.  This is a topic for further work.

{\bf Acknowledgments}

This work was supported in part by DOE grant number DE-FG03-92ER40689. I would like to thank Tom Banks and Michael Dine for many helpful discussions.

\bibliographystyle{apsrev}

\end{document}